# Jones-matrix analysis of phase accumulation in a linear-optical multi-pass interferometer

Byoung S. Ham[1,2]

[1]Department of Electrical Engineering and Computer Science, Gwangju Institute of Science and Technology, 123 Chumdangwagi-ro, Buk-gu, Gwangju 61005, South Korea
[2]Department of Electrical and Computer Engineering, Oregon State University, Corvallis, OR 97331, USA
(June 12, 2026; bham@gist.ac.kr)

**Abstract**
Quantum information science has traditionally relied on nonclassical resources, such as entangled photon pairs and squeezed states, to achieve measurement performance beyond classical limits. Here, we revisit the multi-pass photonic scheme reported in Nature **450**, 393 (2007) to clarify the physical origin of the observed superresolution and the associated claim of supersensitivity. Using a rigorous Jones-matrix formalism, we show that the round-trip evolution of the HQMQ linear optics unit is equivalent to the product of two reflections in polarization space, resulting in an effective rotation operator. This equivalence reveals that the accumulated phase arises from coherent polarization-state rotation on the Poincaré sphere. The resulting phase accumulation is interpreted geometrically as a progressive realignment of the polarization state during successive forward and backward propagations. To validate the theoretical model, a classical-wave implementation is experimentally conducted, analyzed, and compared with the corresponding Jones-matrix solution. Finally, the scaling behavior of the Fisher information is analyzed to examine the origin of the claimed supersensitivity. The results are further compared with a recently developed coherence de Broglie wavelength framework, which achieves identical superresolution through repeated coherent interactions in a cascaded interferometeric architecture.

## 1. Introduction

In quantum information science, the concept of quantum superposition is often interpreted differently from the concept of coherence in classical optics, although both descriptions originate from wave phenomena [1-13]. The distinction is commonly associated with wave-particle duality [3], in which wave-like and particle-like properties provide complementary descriptions of the same physical system [1-3]. In coherence optics, consistent with Maxwell's equations, coherence arises from a well-defined relative phase between optical fields [14]. In practice, however, the coherence of a light source is limited by its finite spectral bandwidth, resulting in a localized wave packet characterized by a finite coherence time and coherence length. This wave-packet description bears a close resemblance to the particle concept represented by Schrodinger's wave function in quantum mechanics. Consequently, the wavelength associated with a quantum particle may be understood as an effective property emerging from an ensemble of monochromatic wave components. Within this framework, the uncertainty relation between conjugate variables, such as position and momentum, or photon number and phase, follows naturally from Fourier analysis. As the uncertainty in photon number decreases, the uncertainty in phase correspondingly increases. In the limiting case of a Fock state with a well-defined photon number, a well-defined absolute optical phase cannot be assigned [4-13]. This observation raises fundamental questions regarding the physical interpretation of phase in quantum measurements and motivates a careful distinction between absolute phase, relative phase, and phase correlations established between multiple particles or optical modes.

Over the last century, quantum mechanics has been largely developed from the particle interpretation of Schrodinger's wave equation, emphasizing measurement outcomes associated with discrete quantum particles [4-13]. Within this framework, the role of phase has often been discussed primarily through probability amplitudes rather than through direct physical phase relations between individual particles. Consequently, quantum phenomena such as nonlocal correlation [4-10], quantum erasers [15], and Hong-Ou-Mandel interference [16] have traditionally been interpreted within the standard quantum-mechanical formalism. More recently, alternative approaches have been explored to reexamine these phenomena from the perspective of optical coherence and phase correlations [17-20]. In these studies, the phase degree of freedom plays a central role in analyzing effects that are otherwise regarded as uniquely quantum. For example, the wave-packet description has been employed to investigate interference fringes in second-order intensity correlations [17,20]. In conventional coherence optics, first-order interference between orthogonal basis states is forbidden [14,19]; however, phase coherence can be recovered through selective projection measurements at the output ports, at the cost of a deterministic 50% reduction in the available measurement resources [19,20]. More recently, a quantum-mechanical analysis has provided a general description of superresolution in coherence de Broglie waves (CBW) [17], where the N00N-like superresolution [8,11-13] originates from the Nth power of the unitary transformation of a Mach-Zehnder interferometer (MZI). In this picture, the phase associated with a single MZI unit is coherently accumulated through repeated unitary operations, leading to an N-fold phase dependence [21]. This mechanism differs conceptually from conventional resolution enhancement techniques employed in optical sensors, such as wavelength meters [22], gyroscopes [23], and LIGO [24], where the improved resolution is typically achieved through geometric scaling [14], amplitude superposition [22] and increased interaction length [23] via resonant enhancement rather than through repeated coherent application of a unitary phase operator [17,21].

Here, we revisit ref. 25 to analyze the origin of the reported superresolution and claimed supersensitivity in a simple multi-pass structure composed entirely of linear optical elements. For the superresolution, a general solution is derived within a coherence-based framework, showing that the observed superresolution originates from the Nth-power of a unitary transformation of the linear optical elements. The associated phase-accumulation mechanism is then examined through the coherent evolution of polarization bases, where the round-trip Jones-matrix analysis is shown to be equivalent to a rotation of the polarization-state vector. For the claimed supersensitivity, classical Fisher information [26] is also calculated and compared with a recently proposed CBW method [17,21]. Particular attention is paid to the role of the scaling parameter used in ref. 25 and to the interpretation of the random variable and resource counting in Fisher information analysis. Through these analyses, the reported phenomena are reexamined from the perspective of optical coherence and wave evolution, providing an alternative interpretation of effects that are often regarded as uniquely quantum. The purpose of this work is not to challenge the validity of quantum mechanics, but rather to provide additional physical insight into the origin of superresolution and supersensitivity. A deeper understanding of these mechanisms may help bridge coherence optics and quantum information science, thereby contributing to the development of scalable quantum technologies, especially in quantum sensing limited by the rapidly decreasing generation efficiency [8,13,27] and unavoidable fringe degradation [28] of higher-order N00N states.

## 2. Results

### *2-1. Outlook*

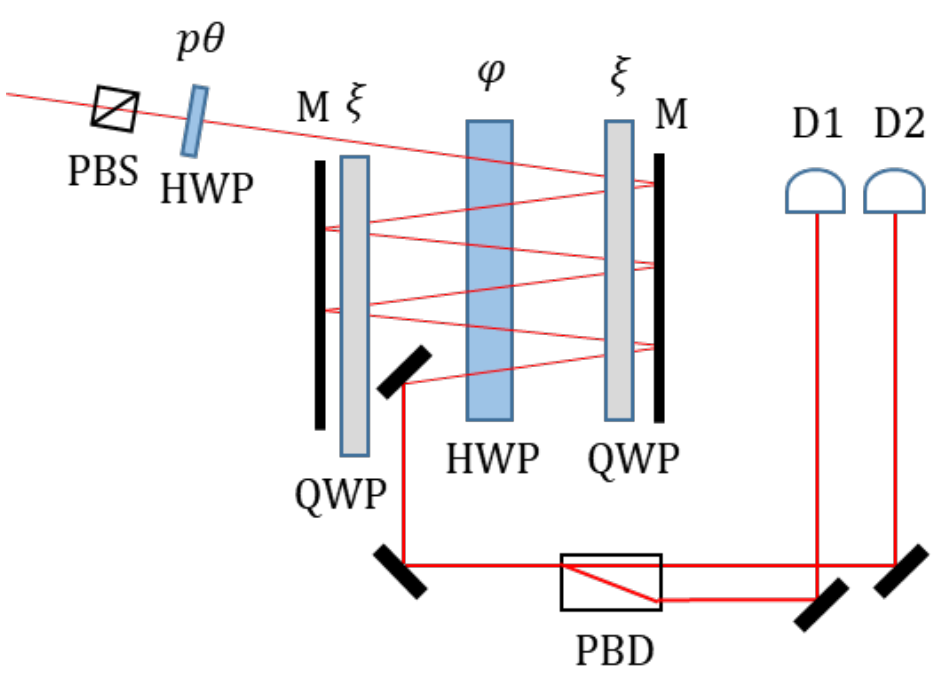

**Fig. 1.** Schematic of entanglement-free Heisenberg-limited phase estimation. D: photodetector; HWP: half-wave plate; M: mirror; PBS: polarizing beam splitter; PBD: polarizing beam displacer; QWP: quarter-wave plate.

Figure 1 illustrates the entanglement-free Heisenberg-limited phase estimation scheme reported in ref. 25. The input light consists of single photons generated from an attenuated laser, whose photon statistics follow a Poisson distribution. A polarizing beam splitter (PBS) prepares the incident photons in the horizontal polarization state. The subsequent half-wave plate HWP($p\theta$) transforms the photons into a coherent superposition state of horizontal and vertical polarization bases: $|\psi_{in}\rangle = \frac{1}{\sqrt{2}}\left(|H\rangle + e^{ip\theta}|V\rangle\right)$. The polarization-superposition state $|\psi_{in}\rangle$ then enters a multi-pass system containing a HWP($\phi$) ($H(\phi)$) enclosed by cavity-like mirrors (Ms). To maintain a multi-pass resulting phase accumulation structure, a quarter-wave plate (QWP(ξ); Q(ξ)) is inserted symmetrically between $H(\phi)$ and the mirror M. The resulting Q-M-Q configuration on both sides of $H(\phi)$ is for realignment of the polarization basis of the photon between successive interactions on $H(\phi)$ (see below). In particular, the reversal of the propagation direction of the photon by the mirror M plays an essential role in the phase accumulation process through the double-pass configuration of $Q(\xi)$. Without $Q(\xi)$, the double-pass $H(\phi)$ scheme cannot satisfy the phase accumulation due simply to a toggle switch-like function with $e^{iN\pi}$ in phase shift. Thus, the unit-phase shift generated by the elementary H-Q-M-Q cell is coherently added during successive passes, yielding the final state $|\psi\rangle_{final} = \frac{1}{\sqrt{2}}\left(|H\rangle + e^{i[N(\phi-\xi)-p\theta]}|V\rangle\right)$. To measure the accumulated phase $N(\phi-\xi)$ in $|\psi\rangle_{final}$, the polarization projection measurement is performed using an additional HWP(ζ), resulting in a coherent superposition of the orthogonal basis components of $|\psi\rangle_{final}$ onto both projection axes (see below). The coherent superposition in the projection process results in interference fringes, $1 \pm \sin[2N(\phi-\xi)]$. Finally, the superimposed out-of-phase fringes are spatially separated by a polarization beam dispenser (PBD), where the measurements are for the first-order intensity correlation.

*2-2. Jones matrix analysis*

The forward-propagating Jones matrices for a HWP and QWP oriented at an arbitrary angle α from the horizontal axis are represented by $J_{HWP}(\alpha) = \begin{pmatrix} cos2\alpha & sin2\alpha \\ sin2\alpha & -cos2\alpha \end{pmatrix}$ and $J_{QWP}(\alpha) = \begin{pmatrix} cos^2\alpha + isin^2\alpha & (1-i)sin\alpha cos\alpha \\ (1-i)sin\alpha cos\alpha & sin^2\alpha + icos^2\alpha \end{pmatrix}$, where $J_{HWP}(\alpha) = R(\alpha)\begin{pmatrix} 1 & 0 \\ 0 & -1 \end{pmatrix}R(-\alpha)$, $J_{QWP}(\alpha) = R(\alpha\xi)\begin{pmatrix} 1 & 0 \\ 0 & i \end{pmatrix}R(-\alpha)$, and $R(\alpha) = \begin{pmatrix} cos\alpha & -sin\alpha \\ sin\alpha & cos\alpha \end{pmatrix}$ is a rotation matrix. A normal-incidence mirror reflection flips the horizontal coordinate frame relative to the direction of propagation, acting as a transformation matrix $J_M = \begin{pmatrix} 1 & 0 \\ 0 & -1 \end{pmatrix}$ [14]. Crucially, when the photon reverses its spatial propagation vector $\mathbf{z} \rightarrow -\mathbf{z}$, any physical orientation angle α transforms to −α relative to the traveling wave's local frame. This reverse feature by M is the key to making the Nth power of $J_{HWP}(\alpha)$, resulting in the N-accumulated phase observed in ref. 25.

*2-3. Evaluation of Leg 1 (right-side loop)*

In Fig. 1, the polarization-superposed photon by HWP($p\theta$) travels to the right, transits the center HWP($\phi$), transits the right QWP($\xi$), reflects off the right mirror M, and returns through the right QWP(−ξ). Grouping the operators from right to left for the unit cell of H-Q-M-Q linear optics chain is represented by $J_{leg1} = J_{QWP}(-\xi)J_M J_{QWP}(\xi)J_{HWP}(\phi)$, where the basis-realignment linear optics is denoted by $J_{QWP}(-\xi)J_M J_{QWP}(\xi) = \begin{pmatrix} cos2\xi & sin2\xi \\ -sin2\xi & cos2\xi \end{pmatrix}$. Thus,

$$J_{leg1}(\phi;\xi) = \begin{pmatrix} cos2\Omega & sin2\Omega \\ sin2\Omega & -cos2\Omega \end{pmatrix}, \quad (1)$$

where $\Omega = \phi - \xi$. Leg 1 acts as a pure geometric reflection matrix $J_{leg1} \equiv J_+ = H(\Omega)$, where $H(\Omega) = H(\phi-\xi) = Q(-\xi)MQ(\xi)H(\phi)$.

*2-4. Evaluation of Leg 2 (left-side loop)*

The photon is now traveling left in an inverted coordinate frame due to the first mirror reflection on the right side. It passes the central HWP($\phi$) (now experiencing an effective angle of $-\phi$), transits the left QWP($-\xi$), reflects off the left mirror M, and returns through the left QWP($\xi$). Evaluating this sequence in the inverted coordinate space yields:

$$J_{leg2}(-\phi;-\xi) = \begin{pmatrix} cos2\Omega & -sin2\Omega \\ -sin2\Omega & -cos2\Omega \end{pmatrix}, \quad (2)$$

where the Jones matrix of the normal backward propagation light is $J_{HWP}(-\phi) = R(-\phi)\begin{pmatrix} 1 & 0 \\ 0 & -1 \end{pmatrix}R(\phi) = \begin{pmatrix} cos2\phi & -sin2\phi \\ -sin2\phi & -cos2\phi \end{pmatrix}$. Leg 2 acts as a pure geometric reflection matrix about the opposite axis: $J_{leg2} \equiv J_- = H(-\Omega)$.

*2-5. Evaluation of one full unit repeat (N=2 bounces)*

The total transformation for one complete round trip across HWP($\phi$) via the symmetric QMQ configuration requires multiplying the operators for Leg 1 and Leg 2:

$$J_{RT}(N=2) = J_{leg2} \cdot J_{leg1} = \begin{pmatrix} cos4\Omega & sin4\Omega \\ -sin4\Omega & cos4\Omega \end{pmatrix} = R(-4\Omega). \tag{3}$$

Equation (3) represents a typical rotational matrix whose rotation angle is a quadruple of $\xi - \phi$, which is twice that of a single chain of H($\phi$) and Q($\xi$). For the round trip, $J_{RT}(N=2) = J_- \cdot J_+ = H(-\Omega)H(\Omega)$ is satisfied, where $H(a)H(\mathrm{b}) = R[2(a-b)]$. Thus, Eq. (3) mathematically proves phase accumulation by the unit cell of linear optics composed of H-Q-M-Q, where the role of the QMQ set is essential for the basis realignment process. Since Eq. (3) is equivalent to a pure rotational matrix, every time it makes a round trip, both phase and geometrical rotation linearly increase.

*2-6. General N-pass output state equations*

Because rotation matrices satisfy the property $\mathrm{R}(\gamma)^m = R(m\gamma)$, cascading this unit process N times leads to linear angular growth. Given an arbitrary linear input angle χ, where $|\psi_{in}\rangle = \begin{pmatrix} cos\chi \\ sin\chi \end{pmatrix}$ and $\chi = p\theta$ in Fig. 1, the final output state vector of the N-passed photon through HWP($\phi$) becomes:

$$|\psi_{out}\rangle = \begin{cases} \begin{pmatrix} \cos(2N\Omega - \chi) \\ sin(2N\Omega - \chi) \end{pmatrix} & if\ N\ is\ odd \\ \begin{pmatrix} \cos(2N\Omega - \chi) \\ -sin(2N\Omega - \chi) \end{pmatrix} & if\ N\ is\ even \end{cases} \tag{4}$$

*2-7. Basis transformation and geometric phase derivation*

To view the accumulated phase shift from the spatial orientation in $|\psi_{out}\rangle$, we map the linearly rotated output vector into the left- and right-circular polarization bases ($|R\rangle, |L\rangle$). For an even number N, let $\Phi_N = 2N\Omega$: $|\psi_{out}\rangle = \begin{pmatrix} \cos(\Phi_N - \chi) \\ -sin(\Phi_N - \chi) \end{pmatrix}$. Applying the circular basis transformation matrix $\mathrm{T} = \frac{1}{\sqrt{2}}\begin{pmatrix} 1 & -i \\ 1 & i \end{pmatrix}$:

$$|\psi_{out}\rangle_{circular} = \frac{1}{\sqrt{2}}\begin{pmatrix} \cos(\Phi_N - \chi) + isin(\Phi_N - \chi) \\ \cos(\Phi_N - \chi) - isin(\Phi_N - \chi) \end{pmatrix} = \frac{1}{\sqrt{2}}\begin{pmatrix} e^{i(\Phi_N - \chi)} \\ e^{-i(\Phi_N - \chi)} \end{pmatrix}. \tag{5}$$

Equation (5) can also be expressed as a state vector: $|\psi_{out}\rangle = \frac{1}{\sqrt{2}}\left(e^{i(\Phi_N-\chi)}|R\rangle + e^{-i(\Phi_N-\chi)}|L\rangle\right)$. This expression proves that the physical rotation of the linear polarization state $|\psi_{in}\rangle$ by N-repeated H-Q-M-Q acts as a non-dispersive phase shifter. From Eq. (5), we can peek at the physical mechanism of phase accumulation by the QMQ unit cell. The right-circular mode picks up a clean, continuous phase advance of $2N(\phi - \xi)$ radians, while the left-circular mode drops by $-2N(\phi - \xi)$ radians:

$$|\psi_{out}\rangle = \frac{1}{2}\left[e^{i(\Phi_N-\chi)}(|H\rangle + i|V\rangle) + e^{-i(\Phi_N-\chi)}(|H\rangle - i|V\rangle)\right]$$

$$= cos(\Phi_N - \chi)|H\rangle - sin(\Phi_N - \chi)|V\rangle. \tag{6}$$

Thus, the probability of the horizontal (vertical) component projected onto a ζ-rotated HWP is $P_H = cos^2(\Phi_N - \chi + 2\zeta) = \frac{1}{2}[1 + cos2(\Phi_N - \chi + 2\zeta)]$ and $P_V = sin^2(\Phi_N - \chi + 2\zeta) = \frac{1}{2}[1 - cos2(\Phi_N - \chi + 2\zeta)]$. If the projection HWP is the same as the control HWP($\phi$), the effective fringe multiplier increases as $P_H = cos^2(N\Phi_N - \chi + 2\phi)$ and $P_V = sin^2(N\Phi_N - \chi + 2\phi)$. The role of PBD is to spatially separate the two orthogonal projection components, $P_H$ and $P_V$, producing complementary out-of-phase superresolution fringes. The N-dependent accumulated angle $\Phi_N = 2N(\phi - \xi)$ is therefore converted into N-fold polarization-fringe superresolution.

*2-8. Quantum mechanical analysis*

Assuming lossless linear optics of HWP, QWP, and mirror M, every optical element is represented by a unitary Jones matrix $\mathrm{U} \in \mathrm{SU}(2)$: $H^\dagger H = Q^\dagger Q = M^\dagger M = I$. Thus, for a single photon, the output photon state also satisfies the unitary relation: $|\psi_{out}\rangle = U|\psi_{in}\rangle$. For the unit cell of linear optics composed of HQMQ in the right propagation direction in Fig. 1, the corresponding unitary operator is $U_{unit} = U_Q U_M U_Q U_H = \begin{pmatrix} cos2\Omega & sin2\Omega \\ -sin2\Omega & cos2\Omega \end{pmatrix}$, where $\Omega = \phi - \xi$. This HQMQ is for the first half of a complete round trip. This unit operator $U_{unit}$ is a real rotation matrix, $R(-2\Omega)$. Here, it should be noted that $U_H$ is not a rotation matrix:$U_H U_H = I$. With the opposite propagation direction in the second half, the mirror-symmetric counterpart of the second HQMQ results in the same unitary matrix as $U_{unit}$. Thus, the round-trip unitary matrix is $U_{RT} = U_{unit} U_{unit} = \begin{pmatrix} cos4\Omega & sin4\Omega \\ -sin4\Omega & cos4\Omega \end{pmatrix}$. This is again a real rotation matrix but with a doubled phase, $R(-4\Omega)$. As the unit cycle is repeated by N times, the unit phase is linearly accumulated to be $2N\Omega$ by the action of $(U_{unit})^{\otimes N}$ to the output photon state: $|\psi_{out}\rangle = (U_{unit})^{\otimes N}|\psi_{in}\rangle$. This phase accumulation corresponds to geometrical rotation of the Stokes vector on the Poincaré sphere (see Discussion). At this point, one may ask what if the accumulated rotation angle is greater than $2\pi$? Although the geometric rotation cannot be differentiated from the unit rotation of $2\pi$, the accumulated phase $2N\Omega$ in $|\psi_{out}\rangle$ can be extracted via polarization projection measurements, as analyzed above.

The polarization-basis projection mechanism through $H(\zeta = 22.5°)$ is equivalent to the Hadamard transformation: $|H\rangle \rightarrow \frac{|H\rangle+|V\rangle}{\sqrt{2}}$ and $|V\rangle \rightarrow \frac{|H\rangle-|V\rangle}{\sqrt{2}}$. For the final state $|\psi_{final}\rangle = \frac{1}{\sqrt{2}}(|H\rangle + e^{i2N\Omega}|V\rangle)$, the Hadamard transformation results in $|\psi_{out}\rangle = \frac{1}{2}[(1 + e^{i2N\Omega})|H\rangle + (1 - e^{i2N\Omega})|V\rangle]$. Thus, $P_H = \frac{1}{2}\left|1 + e^{i2N\Omega}\right|^2 = \frac{1}{2}[1 + \cos(2N\Omega)]$ and $P_V = \frac{1}{2}\left|1 - e^{i2N\Omega}\right|^2 = \frac{1}{2}[1 - \cos(2N\Omega)]$ are resulted.

## 3. Experiments

Figure 2 represents an experimental demonstration of the superresolution mechanism analyzed above in a pure classical regime. Unlike the single-photon implementation of ref. 25, CW laser light is employed here to highlight the classical origin of the observed superresolution. As a proof-of-principle experiment, the N=2 case is compared directly with the N=1 case (reference). The upper panel illustrates the N=2 configuration, in which the photon undergoes a double pass through $H(\phi)$ via a round-trip propagation enabled by the $Q(\xi)M$ assembly, resulting in a HQMQ-HQMQ configuration. The output field is subsequently passed through $H(\zeta)$, which performs a polarization-projection measurement by transforming the accumulated phase into a polarization-basis superposition. The projected field is then analyzed by a polarization beam displacer (PBD), which spatially separates the horizontal (H) and vertical (V) polarization components. These H/V components correspond to complementary projection bases of the output field and exhibit interference fringes that are mutually out of phase. The separated beams are projected onto a screen and recorded using a camera (see the video files). The counting number of fringes per unit scanning of $H(\phi)$ is twice of the reference (see https://studio.youtube.com/video/4GIH21Tz8aU/edit): To generate the interference patterns, $H(\phi)$ is continuously rotated, thereby scanning the phase parameter $\phi$.

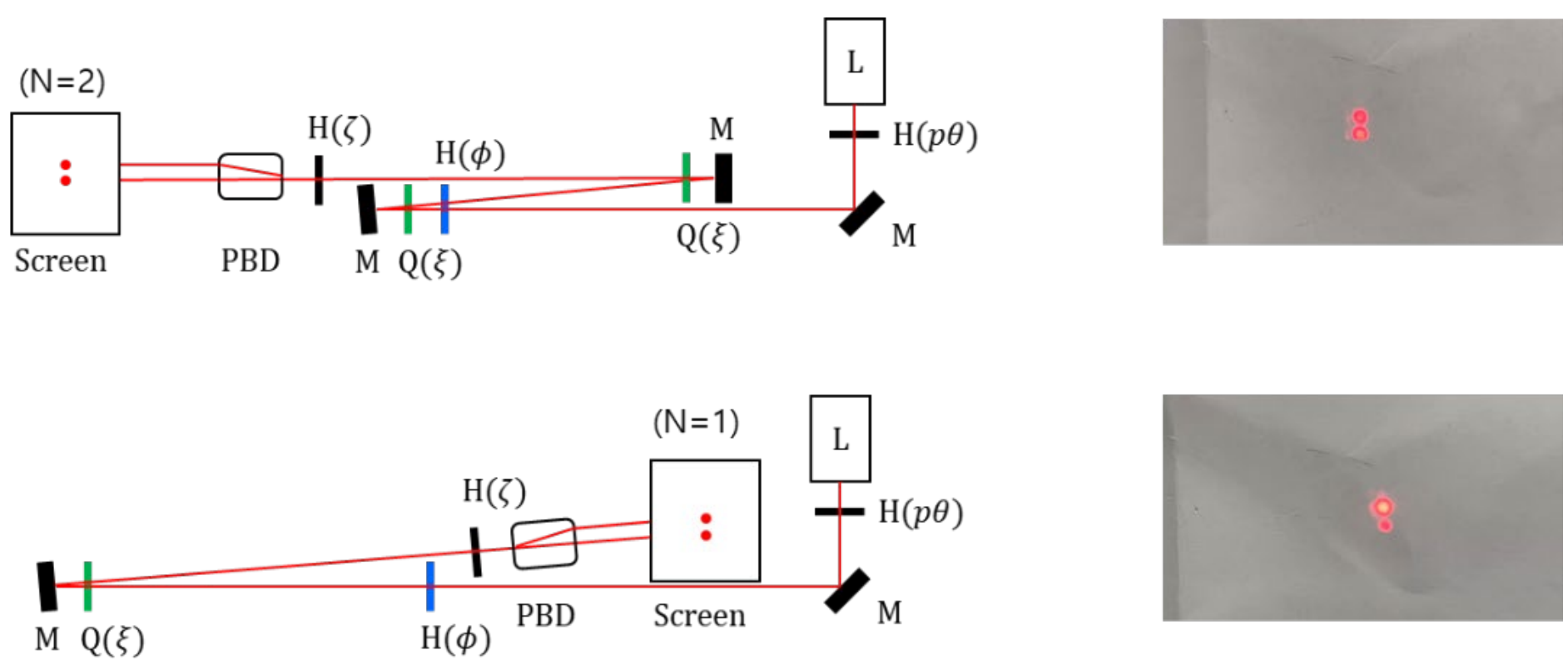

**Fig. 2.** Experimental demonstration of multi-pass superresolution using a CW input light. PBD: polarization beam dispenser. H: Half-wave plate, M: mirror, Q: quarter-wave plate. Upper panel: N=1 (https://studio.youtube.com/video/nTsm_oR2LCY/edit). Lower panel: N=2 (see https://studio.youtube.com/video/4GIH21Tz8aU/edit).

For comparison, the lower panel shows the corresponding single-pass (N=1) configuration. Although the scanning speeds of $H(\phi)$ for the N=1 and N=2 measurements are not perfectly identical because of manual operation, they were maintained at approximately the same rate. As shown in the recorded video sequences, the N=2 configuration exhibits a clear doubling of the fringe frequency compared with the N=1 case, in agreement with the theoretical prediction of phase accumulation. When the output field is observed without PBD, no interference fringes are visible because the two complementary polarization projections are superimposed, causing the out-of-phase fringe patterns to average out (not shown).

Importantly, the observed superresolution does not depend on the quantum nature of the input light. According to the standard interpretation of quantum mechanics, interference arises from the coherent evolution of individual probability amplitudes rather than from interactions between photons. Consequently, the phase-accumulation mechanism is expected for both single-photon and classical CW light, where the Fisher information for both cases is equivalent. The experimentally observed superresolution in ref. 25, therefore, originates not from any nonclassical property of the photons, but from the coherent polarization-basis realignment and polarization-state evolution induced by the linear-optical QMQ module during repeated passes through the HQMQ system.

## 4. Discussion

### *4-1. Geometrical interpretation*

The superresolution coherently derived in Eq. (6) and observed in ref. 25 is now discussed in terms of basis realignment through-H-Q-M-Q unit-cell operations. For a linearly polarized input state, $|\psi\rangle_{in} = \frac{1}{\sqrt{2}}(cos\beta|H\rangle + sin\beta|V\rangle)$, the corresponding Jones vector in the H/V basis is $|\psi\rangle_{in} = \begin{pmatrix} cos\beta \\ sin\beta \end{pmatrix}$, where $\beta = p\theta$. On the Poincaré sphere, this state is represented by the Stokes vector $S = (S_1, S_2, S_3)$, where $S_1 = \frac{I_H - I_V}{I}$, $S_2 = \frac{I_D - I_A}{I}$, and $S_3 = \frac{I_R - I_L}{I}$. Thus, the tip of the state vector lies on the equator with $|H\rangle \rightarrow (1,0,0)$ and $|V\rangle \rightarrow (-1,0,0)$. After passing through the $H(\phi)$, the polarization angle is transformed as $\gamma = 2\phi - \beta$. Thus, the field entering the first $Q(\xi)$ is represented as $|\psi\rangle_H = \begin{pmatrix} \cos\gamma \\ sin\gamma \end{pmatrix} = cos\gamma|H\rangle + sin\gamma|V\rangle$. Using the circular-basis definition of $|R\rangle = \frac{|H\rangle - i|V\rangle}{\sqrt{2}}$ and $|L\rangle = \frac{|H\rangle + i|V\rangle}{\sqrt{2}}$, this state can be rewritten as $|\psi\rangle_H = \frac{1}{\sqrt{2}}(e^{i\gamma}|R\rangle + e^{-i\gamma}|L\rangle)$. Hence, the linearly polarized field entering the first $Q(\xi)$ can equivalently be viewed as a balanced superposition of right- and left-circular components with opposite phases $\pm\gamma$.

### *4-2. A single photon vs continuous-wave light*

In ref. 25, the role of a single photon has been emphasized in explaining the quantum nature of the observed superresolution. In addition, the supreresolution-associated supersensitivity has been claimed by the authors. Having established the Jones-matrix and Jones-vector description of the repeated HQMQ process, the remaining question concerns the physical interpretation of the input field. As shown by the projection measurement of Eq. (6), the observed superresolution originates from coherent polarization-state evolution and the subsequent interference between orthogonal polarization components. The resulting fringes are mathematically analogous to those observed in a conventional Mach-Zehnder interferometer (MZI), where interference arises from the coherent superposition of two optical paths. In the present system, the interference occurs between two orthogonal polarization modes rather than between spatially separated paths.

From this perspective, the formation of the superresolution fringe pattern observed in ref. 25 is governed by coherence and basis projection, as described by Eq. (6). Consequently, the N-fold fringe multiplication derived above does not depend on the photon-number character of the input field (see Fig. 2). The same interference pattern can be obtained with a single-photon state, coherent CW light, or other optical states, provided that the required coherence between the polarization components is maintained. The distinction between these input states becomes important when evaluating photon statistics, noise properties, and phase-estimation sensitivity, rather than in the coherent phase-accumulation mechanism itself. Furthermore, the single-photon measurements reported in ref. 25 were performed by detecting individual photon events over many experimental trials. Thus, the observed interference fringes arise from the accumulated statistics of repeated measurements and are fully described by the coherent evolution and projection process derived above.

*4-3. Super-resolution vs. super-sensitivity*

In ref. 25, the authors claimed supersensitivity based on the observed superresolution fringes of the form $1 \pm \cos(N\phi)$. Although the N-fold phase dependence clearly demonstrates superresolution, the interpretation of the corresponding phase sensitivity requires careful consideration. In estimation theory, Fisher information is defined through the probability distribution of measurement outcomes conditioned on an unknown parameter. The resulting Cramer-Rao lower bound (CRLB) therefore depends on the statistics of repeated measurement events rather than on the functional form of the fringe alone.

For the multi-pass system considered in ref. 25, the parameter N denotes the number of coherent round trips experienced by a single photon during a single measurement event. Thus, N is a fixed geometric parameter of the interferometric structure rather than a stochastic resource associated with repeated independent measurements, say M. Although the phase response is amplified from $\phi$ to $N\phi$, yielding fringes of the form $1 \pm \cos(N\phi)$, this amplification by itself does not establish Heisenberg-limited scaling due to $N \neq M$. A meaningful comparison with the shot-noise limit (SNL) or the Heisenberg limit requires explicit accounting of the total measurement resources and the statistical ensemble used to estimate the unknown phase.

From the coherent analysis presented above, the parameter N originates from repeated basis realignment during the multi-pass evolution. As shown in Eq. (6), each round trip contributes to an identical unitary rotation, resulting in an accumulated phase proportional to N. Consequently, N serves as an enhancement parameter associated with the geometry of the multi-pass system. The same phase-accumulation mechanism has recently been demonstrated in a cascade MZI, where the basis realignment occurs between path states rather than polarization states [29]. In both systems, the accumulated phase arises from repeated coherent evolution accompanied by basis realignment. Without such a basis realignment, successive round-trip phases simply undo one another, resulting in no net phase accumulation.

*4-4. Role of super-resolution*

By the first-pass $Q(\xi)$ in Fig. 1, the linearly polarized state $|\psi\rangle_H$ is transformed into $|\psi\rangle_{Q1} = \frac{1}{\sqrt{2}}\left(e^{i(\gamma-\xi)}|R\rangle + e^{-i(\gamma-\xi)}|L\rangle\right)$, where $|R\rangle$ and $|L\rangle$ denote the right- and left-circular polarization bases, respectively. Rather than generating a purely circularly polarized state, the QWP introduces a relative phase shift between the two circular polarization components. The subsequent mirror M reverses the handedness of circular polarization $|R\rangle \leftrightarrow |L\rangle$, so that the photon state after reflection becomes $|\psi\rangle_M = \frac{1}{\sqrt{2}}\left(e^{-i(\gamma-\xi)}|R\rangle + e^{i(\gamma-\xi)}|L\rangle\right)$. Passing through the second $Q(-\xi)$, the state is transformed into $|\psi\rangle_{out} = \frac{1}{\sqrt{2}}(e^{-i\gamma}|R\rangle + e^{i\gamma}|L\rangle)$. Expressing this state in the linear polarization basis yields $|\psi\rangle_{out} = cos\gamma|H\rangle - sin\gamma|V\rangle$. Comparing this result with the input state to the QMQ section, $cos\gamma|H\rangle + sin\gamma|V\rangle$, shows that the net action of the QMQ sequence is equivalent to $(H,V) \rightarrow (H,-V)$ or, in Jones-matrix form, $\sigma_z = \begin{pmatrix} 1 & 0 \\ 0 & -1 \end{pmatrix}$.

On the Poincaré sphere, this transformation corresponds to $(S_1, S_2, S_3) \rightarrow (S_1, -S_2, -S_3)$. For a linear polarization ($S_3 = 0$), the polarization angle after the QMQ section becomes $\beta_{out} = 2\xi - (2\alpha - \beta) = 2(\xi - \alpha) + \beta$, where $\xi$ and $\alpha$ are the physical (geometric) rotation angles of the QWP and HWP, respectively. Thus, the QMQ section performs a basis realignment, preparing the output state so that it acts as the appropriate input state for the next $HWP(\phi)$ interaction. Consequently, successive HQMQ operations coherently add the same angular increment $\beta_{N+1} = \beta_N + 2(\xi - \alpha)$, which yields $\beta_N = 2N(\xi - \alpha) + \beta$. This result is equivalent to the repeated application of the unit-cell rotation operator $R(-4\Omega)$, where $\Omega = \alpha - \xi$. Therefore, after N repetitions, the accumulated polarization rotation angle is $-2N\Omega$.

The Stokes vector on the Poincaré sphere remains on the equator and evolves according to $(cos2\beta_n, sin2\beta_n, 0)$, corresponding to a uniform angular walk generated by repeated basis realignment. Physically, the QMQ section converts the polarization state into circular-polarization eigenmodes, allows the mirror to reverse its handedness, and then converts them back into the linear basis. The resulting basis realignment prevents successive round trips from cancelling one another and instead enables coherent accumulation of the polarization angle. Consequently, the phase accumulation $2N(\xi - \alpha)$ is a direct manifestation of repeated coherent evolution under linear optics. The accumulated phase is finally extracted through polarization-basis projection, where interference between orthogonal polarization components produces the N-fold superresolution fringes derived in Eq. (6).

## 5. Conclusion

In conclusion, the Jones-matrix analysis of the round-trip photon configuration revealed that the phase-accumulation mechanism in the multi-pass HQMQ system could be fully understood within the framework of coherent polarization-state transformations. For a single half-cycle, the operator QMQ reduced to an effective reflection matrix $H(\phi - \xi)$, indicating that the polarization state was reflected about an axis determined by the relative orientation of the HWP and QWP. The backward-propagating half-cycle similarly yielded the complementary reflection ($H[-(\phi - \xi)]$. The complete round-trip was therefore represented by the product of two reflections, which is mathematically equivalent to a rotation operator ($R[4(\xi - \phi)]$). As a result, repeated round trips generated a uniform rotation of the polarization-state (Stokes) vector along the equator of the Poincaré sphere, producing a cumulative phase that increases linearly with the number of cycles. This analysis established the equivalence between the explicit Jones-matrix description of the forward-backward photon evolution and the geometrical interpretation of the HQMQ unit cell as an effective polarization-rotation operator. The observed superresolution originated from this coherent accumulation of polarization rotation. Since each complete cycle contributed an identical angular increment, the accumulated phase after N cycles scaled as $N(\phi - \xi)$, resulting in an interference signal with N-fold enhancement in phase resolution. To support this theoretical (mathematical) derivation and physical interpretation of the phase-accumulation process, proof-of-principle experiments using classical CW laser light were conducted, demonstrating superresolution for N=2. The claim of supersensitivity was also examined from the perspective of estimation theory. Although the accumulated phase contains an N-dependent scaling factor, this parameter is a fixed property of the optical configuration rather than a random measurement variable. Consequently, the observed superresolution arose from coherent phase accumulation in a linear-optical system, while any assessment of sensitivity enhancement must be based on the Fisher information associated with the actual measurement outcomes and the total physical resources employed.

**Acknowledgment**

The author gratefully acknowledges Prof. Ben Lee at Oregon State University for hosting the author during the sabbatical leave and for kindly providing office space in the Department of Electrical and Computer Engineering.

**Data availability**

All data generated or analyzed during this study are included in the published article.

**Author contribution**

B.S.H. conceived the idea and solely wrote the paper.

**Funding**

The author acknowledges that this work was supported by the IITP-ITRC grant (IITP 2026-RS-2021-II211810) funded by the Korean government (Ministry of Science and ICT). BSH also thanks the financial support from GIST Research Project grant funded by the GIST 2026.